# INTEGRAL – status of the mission

**Christoph Winkler[1]**
*Research and Science Support Department, Astrophysics and Fundamental Missions Division*
*ESA-ESTEC, Keplerlaan 1, NL-2201 AZ Noordwijk, The Netherlands*
*E-mail:* `cwinkler@rssd.esa.int`

The ESA gamma-ray observatory INTEGRAL, launched on 17 October 2002, continues to produce a wealth of discoveries and new results on compact high energy Galactic objects, nuclear gamma-ray line emission, diffuse line and continuum emission, cosmic background radiation, AGN, high energy transients and sky surveys. The mission's technical status is healthy and INTEGRAL is continuing its scientific operations well beyond its 5-year technical design lifetime until, at least, 31 December 2014. This paper describes the current status of INTEGRAL including the spacecraft technical state-of-health and the scientific observing programme including its on-going and future multi-year "legacy" programmes..

[1] Speaker

# 1. The INTEGRAL mission

The ESA observatory INTEGRAL [1] is dedicated to the fine spectroscopy (2.5 keV FWHM @ 1 MeV) and fine imaging (angular resolution: 12 arcmin FWHM) of celestial gamma-ray sources in the energy range 15 keV to 10 MeV with concurrent source monitoring in the X-ray (3-35 keV) and optical (V-band, 550 nm) bands. INTEGRAL, with a total launch mass of about 4 tons, was launched on 17 October 2002 from the Baikonur Cosmodrome (Kazakhstan) using a PROTON rocket equipped with a Block DM $4^{th}$ stage. The orbit is characterized by a high perigee in order to provide long periods of uninterrupted observations with nearly constant background and away from trapped radiation (electron and proton radiation belts). The orbital parameters at the beginning of the mission were: 72-hour orbit with an inclination of 52.2 degrees, a height of perigee of 9000 km and a height of apogee of 154000 km.

Owing to background radiation effects in the high-energy detectors, scientific observations are carried out while the satellite is above a nominal altitude of typically 40000 to 60000 km. This means, that most of the time spent in the orbit provided by the PROTON launcher can be used for scientific observations, about 210 ks per revolution. An on-board particle radiation monitor allows the continuous assessment of the radiation environment local to the spacecraft. INTEGRAL carries two main gamma-ray instruments, the spectrometer SPI [2], optimized for the high-resolution (2.5 keV FWHM @ 1 MeV) gamma-ray line spectroscopy (20 keV - 8 MeV), and the imager IBIS [3], optimized for high angular resolution (12 arcmin FWHM) imaging (15 keV - 10 MeV). Two monitors, JEM-X [4] in the (3-35) keV X-ray band, and OMC [5] in the optical Johnson V-band complement the payload. All instruments are co-aligned with overlapping fully coded field-of-views ranging from 4.8º diameter (JEM-X), 5ºx5º (OMC), to 9ºx9º (IBIS) and 16º corner-to-corner (SPI), and they are operated simultaneously, providing the observer with data from all four instruments. The Mission Operations Centre at ESOC (Darmstadt/Germany) performs all standard spacecraft and payload operations and maintenance tasks. The INTEGRAL Science Operations Centre (ISOC) [6] in Madrid (Spain) is responsible for the science operations planning including the implementation of Target of Opportunity observations within the pre-planned observing programme. The INTEGRAL Science Data Centre (ISDC) in Versoix (Switzerland) receives the science telemetry for near-real time monitoring, standard science analysis and archive ingestion [7].

## 1.1 Programmatic and technical status

The nominal 2-year mission operations phase was completed on 1 January 2005, and INTEGRAL is currently being operated in the extended mission phase. Several mission extension requests have been granted since then, and INTEGRAL is currently funded until 31 December 2012, with mission operations extended until 31 December 2014, subject to the usual technical and scientific mid-term review in Fall 2012 to confirm the requested budget beyond 2012. The INTEGRAL spacecraft continues to operate flawlessly. The payload is in good shape after 9 years in orbit. The percentage of healthy detectors is 79% (SPI), 96% (IBIS-ISGRI), 98% (IBIS-PICsIT) and 77% (JEM-X), corresponding to a change in sensitivity compared to its value at launch of 12%, 2%, 1%, 13%, respectively. As of 10 October 2010 (revolution 976), both JEM-X units are being operated simultaneously again.

During 9 years since launch, about 60 kg of on-board fuel, which is used for attitude control and orbit maintenance, have been consumed and about 120 kg of fuel are still available: an average consumption of about 600 g/month. Likewise, the solar array power margin is very comfortable: the spacecraft is still being operated in its nominal post-launch configuration which, thanks to it's solar aspect angle constraint of ±40°, allows a good sky visibility at any point in time, until at least 2014 - 2016. Only then, the degradation of the solar arrays is expected to reduce the array's power output such that the solar aspect angle needs to be constrained to ±30°.

Solar and lunar gravitation are influencing the orbital parameters. For example, the perigee height evolved from 9000 km at the start of the mission to about 13000 km in 2007 and has reached a minimum of 2756 km on 25 October 2011, after which it will increase again to a local maximum of 10000 km in 2016 with subsequent decrease. As of early 2010, the solar activity is picking up towards the next solar maximum in 2013/2014. The corresponding cosmic-ray induced background, being anti-correlated with the solar activity, shows a decreasing background rate in all detectors and veto subsystems since early 2010.

## 2. The observing programme

INTEGRAL is being operated as an observatory. The observing programme, 100% open to the scientific community at large, is built from observing proposals, which have been peer reviewed by the Time Allocation Committee (TAC). Submission of proposals is invited following an Announcement of Opportunity (AO), issued by ESA once a year. In the early years of the mission, the observing programme contained a guaranteed programme (Core Programme) providing science return to the PI teams and to other members of the Science Working Team (Fig. 1). In 2007 (AO-4), ESA introduced a long observation of the Galactic Centre as a pilot "key programme" to the community. The concept of key (legacy) programmes turned out to be very successful and tailor-made for INTEGRAL: a key programme should be at least as long as 1 Ms, and can span two years (two AO-cycles) if justified by its science objectives[2]. Figure 1 shows that the vast majority of open time observations during the past years consists of key programmes. All observations, with the exception of ToO follow-up observations, are open to the scientific community for the submission of so-called "data rights proposals", to obtain data rights on sources which are contained in the large FOV of accepted observations.

---

[2] More information on long-term legacy goals for INTEGRAL observations can be found on the INTEGRAL WWW-site http://www.rssd.esa.int/index.php?project=INTEGRAL&page=IUG in the documents related to the recent (2010) mission extension request.

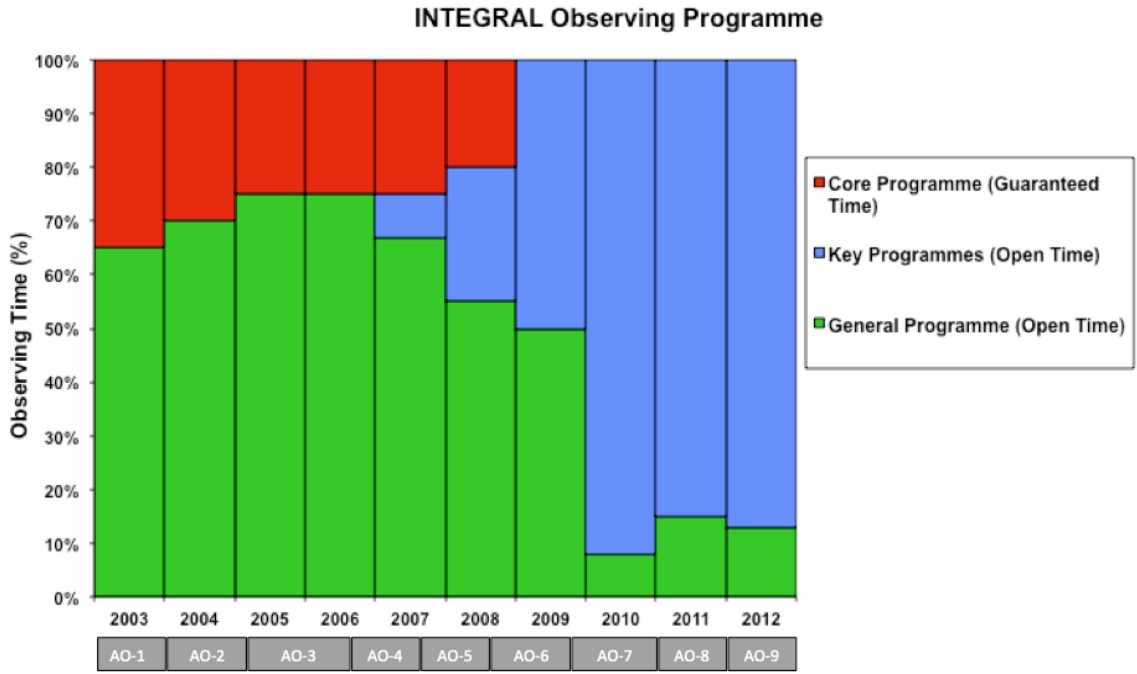

*Figure 1:* The INTEGRAL observing programme. As of 2009, the programme is 100% open to the scientific community, and it is being dominated by large key programme observations.

The accepted observing programme for AO-9, with observations to be scheduled from 1 January 2012 until 31 December 2012 shows, that about 21 Ms out of a total available observing time of about 24 Ms will be devoted to observations lasting 1 Ms or more (Figure 2).

| Science Objectives | MY | Exposure (Ms), PI |
|---|---|---|
| LMC/SN 1987A | MY – new | 3.0, Grebenev |
| Diffuse 511 keV emission: galactic longitude distribution | | 2.5, Weidenspointner |
| Ultra-lum X-ray src/low lum AGN (M82 X-1, Hol X, M81) | MY – new | 2.0, Sazonov |
| Galactic plane scans (monitoring transient sources) | MY – confirmation | 2.0, Bazzano |
| Galactic centre survey | MY – new | 2.0, Wilms |
| Galactic anti-centre survey (l = 150°) | | 2.0, Ubertini |
| Galactic halo: diffuse emission | MY – confirmation | 1.6, Strong |
| 3C 273 region | MY – confirmation | 1.5, Walter |
| Monitoring: inner spiral arms (Per/Nor, Scu/Sgr) | | 1.2, Bodaghee |
| Microquasars GRS 1915+105, Cyg X-1 | MY – confirmation | 1.2, Wilms |
| Galactic diff emission: Carina arm (scans in b at l = -78.5°) | | 1.0, Krivonos |
| Galactic diff emission: GC (scans in b at l = 0°) | | 1.0, Sunyaev |

*Figure 2:* Non-ToO observations as accepted for the AO-9 cycle of INTEGRAL observations (1 Jan – 31 Dec 2012). MY = (multi-year) key programmes spanning 2 AO-cycles (years). <u>New</u> MY programmes will start in AO-9 and can be re-confirmed for AO-10, while <u>confirmed</u> MY programmes have been started in AO-8 and are confirmed for AO-9.

The routine scheduling of accepted open time observations during a calendar year (e.g., Aug 2010 – Aug 2011) shows, that the typical duration of observations ranges between 500 ks and >2 Ms. Shorter exposures (< 500 ks) are performed mostly for ToO follow-up observations. About 20% of the exposure time for non-ToO observations has been used for observations which use scans or customized dither patterns, other than the standard 5x5 or hexagonal dither pointing patterns. About one third of all accepted ToO observations during this time interval were based on unsolicited notifications of new and unforeseen ToO events, i.e. outside the TAC review process of AO-based proposals.

With the AO-9 cycle of observations ahead, the total science observing time, which has been used with INTEGRAL over the time period from 1 January 2003 until 31 December 2012 amounts to about 240 Ms. The total exposure map over this period is shown in Figure 3, in which scheduled observations contribute from 1 January 2003 until November 2011, and planned observations are included according to the long-term plan thereafter, until 31 December 2012.

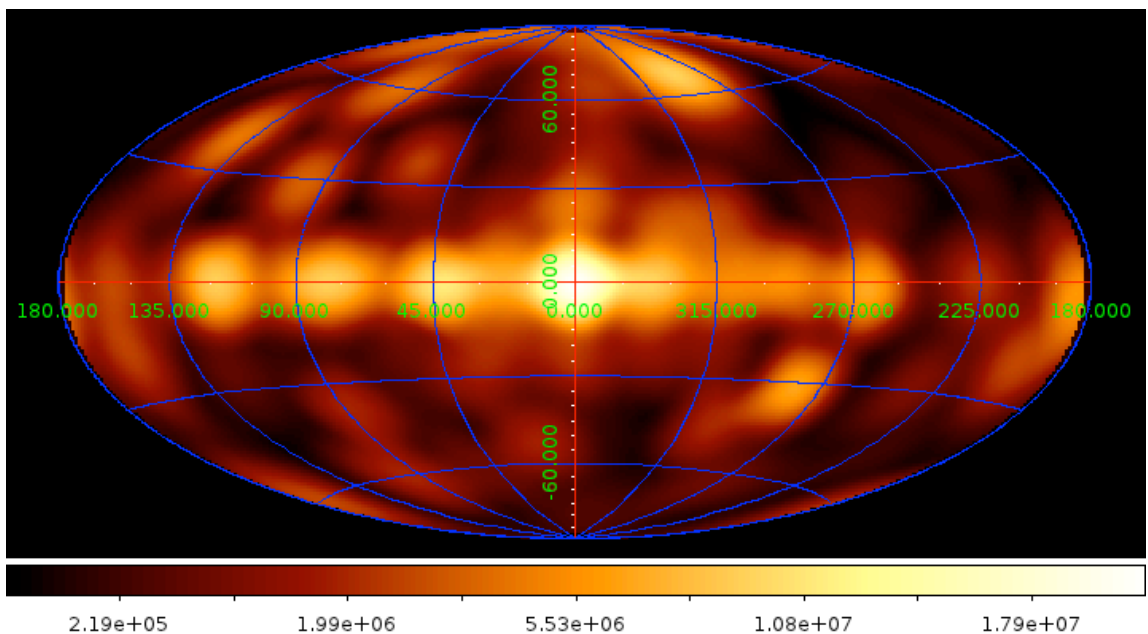

*Figure 3:* *The total exposure map for scheduled and planned observations (Jan 2003 – Dec 2012, AO-1 until AO-9, see text) in galactic co-ordinates, color-scale units are seconds. © ESA/ISOC (G. Bélanger).*

## 3. Science community interfaces

How can the science community interface with the INTEGRAL mission, besides participating in the observing programme as described above?

1. Scientists can access public and private[3] data from various archives: (i) The ISDC is providing consolidated and near-real time data. (ii) Archival data are available at ISDC; at the ISOC; at the INTEGRAL Guest Observer Facility at NASA/GSFC; and at the Russian Science Data Centre at IKI (Moscow). (iii) Instantaneous public data are available e.g. from the SPI anti-coincidence subsystem (light-curves from off-axis GRB and SGR); from Galactic Bulge and Plane monitoring programs; from GRB detected in the FOV (e.g. light-curve, fluence); from unique ToO observations like the SN 2011fe in M101 obtained during October and November 2011; from in-flight calibration observations etc.
2. Target of Opportunity (ToO) observations which provide important serendipitous science, are possible via the execution of TAC approved ToO proposals, which will be scheduled once their specific scientific and operational criteria have been fulfilled. However, in case of a new (unexpected) event, scientists can always alert the INTEGRAL science operations team via a ToO alert web-site in order to consider an observation which is timely and of scientific relevance, even if it is not included in the database of accepted observing programmes of the on-going AO cycle.
3. The scientific users community at large can interface with the INTEGRAL project via membership in two important international committees[4]. Membership is usually for a few years on a rotational basis: (i) The INTEGRAL Time Allocation Committee, which is in charge of peer reviewing all observing and data rights proposals. All submitted proposals would be reviewed on their scientific merit. Successful proposals will be recommended to ESA for implementation. (ii) The INTEGRAL Users Group, which was established by ESA in 2005 as a merger with the original INTEGRAL Science Working Team. Its prime objectives are to: (1) maximize the scientific return of INTEGRAL; (2) ensure, that the INTEGRAL observatory is satisfying the objectives of the scientific community at large, and (3) act as a focus for the interests of the scientific community in INTEGRAL and act as an advocate for INTEGRAL within that community.

**References**

[1] C. Winkler et al., 2003, A&A 411, L1

[2] G. Vedrenne et al., 2003, A&A 411, L63

[3] P. Ubertini et al., 2003, A&A 411, L131

[4] N. Lund et al., 2003, A&A 411, L231

[5] J.M. Mas-Hesse et al., 2003, A&A 411, L261

[6] R. Much et al., 2003, A&A 411, L49

[7] T.J.-L. Courvoisier et al., 2003, A&A 411, L53

---

[3] Private data will become public one year after they have been processed by the ISDC and provided to the PI.
[4] See for more information: http://www.rssd.esa.int/index.php?project=INTEGRAL&page=Teams